\documentclass[pra, reprint, showpacs]{revtex4-1}

\usepackage{amsmath, amssymb}
\usepackage{times, txfonts}
\usepackage[utf8]{inputenc}
\usepackage[english]{babel}
\usepackage{graphicx}
\usepackage{hyperref}

\newcommand{\pllin} {\rho_{({\scriptscriptstyle \rightarrow},
    {\scriptscriptstyle \uparrow})}}

\newcommand{\pldiag} {\rho_{({\scriptscriptstyle \nearrow},
    {\scriptscriptstyle \searrow})}}

\newcommand{\sgn}{\mathop{\mathrm{sgn}}}
\renewcommand{\Im}{\mathop{\mathrm{Im}}}

\begin{document}

\date{November 2, 2016}

\title{Towards spin turbulence of light: Spontaneous disorder and
  chaos in cavity-polariton systems}

\author{S.~S.~Gavrilov}

\affiliation{Institute of Solid State Physics, RAS, Chernogolovka,
  Russia}

\begin{abstract}
  Recent advances in nanophotonics have brought about coherent light
  sources with chaotic circular polarization; a low-dimensional
  chaotic evolution of optical spin was evidenced in laser diodes.
  Here we propose a mechanism that gives rise to light with a
  spatiotemporal spin chaos resembling turbulent states in
  hydrodynamics.  The spin-chaotic radiation is emitted by exciton
  polaritons under resonant optical pumping in arbitrarily sized
  planar microcavities, including, as a limiting case, pointlike
  systems with only three degrees of freedom.  The underlying
  mechanism originates in the interplay between spin symmetry
  breakdown and scattering into pairs of Bogolyubov excitations.  As a
  practical matter, it opens up the way for spin modulation of light
  on the scale of picoseconds and micrometers.
\end{abstract}


\maketitle

\section{Introduction}
\label{sec:introduction}

Deterministic spin chaos has been recently detected in laser
diodes~\cite{Virte13}, promising impressive applications in the field
of information encoding and transmission, fast random number
generation, cryptography, etc.~\cite{Sciamanna15} Today, however, only
a low-dimensional chaotic evolution of optical spin is achieved; also
known are externally shaped~\cite{Colas15} as well as spontaneous
\cite{Tosi12-ncomms} spin patterns.  In this work we address the
problem of making spin of light reveal a \emph{spatiotemporal} chaos
(\cite{Cross93}), similar to turbulent fluids with high-dimensional
chaotic attractors.  We have found that light with such properties can
be emitted by the system of cavity polaritons, half-light half-matter
excitations in microcavities.

Cavity polaritons are composite bosons originating due to the strong
coupling of excitons (electron-hole pairs in semiconductors) and
cavity photons~\cite{Weisbuch92,Yamamoto-book-2000,Kavokin-book-07}.
They are known to form macroscopically coherent states in two ways.
The first is Bose-Einstein condensation, occurring as a phase
transition under nonresonant optical
pumping~\cite{Kasprzak06,Balili07}.  The other way is direct resonant
and coherent driving that immediately governs both density and phase
of a \emph{highly nonequilibrium} polariton
condensate~\cite{Elesin73,Haug83}.  Since excitons are strongly
coupled to externally driven cavity photons, they share a unified
quantum state rather than just constitute a nonlinear medium for
propagating light.

It has long been accepted that chaotic polariton states are feasible
only in inhomogeneous systems and appear due to the Josephson effect.
Josephson oscillations in superconducting junctions occur between
coupled coherent modes whose phase difference varies with
time~\cite{Josephson62}; analogous phenomena were observed in
Bose-Einstein condensates of cold atoms~\cite{Levi07} and cavity
polaritons~\cite{Lagoudakis10,Abbarchi13}.  The Josephson effect is,
however, hardly possible in spatially homogeneous condensates formed
under resonant plane-wave driving.  Polaritons have very small
lifetime, hence a constant harmonic pump force makes them oscillate at
exactly the same frequency, similar to a usual damped pendulum.  Since
the spin-up and spin-down components have constant phase difference,
their interchange---often referred to as intrinsic Josephson
effect~\cite{Shelykh08-j}---is also prevented.  Such a system can only
be \emph{multistable}~\cite{Gippius07, Gavrilov10-en} because of the
polariton-polariton interaction.  Switches between distinct
steady-state branches reveal themselves as sharp jumps in the cavity
field \cite{Baas04-pra, Gippius04-epl, Paraiso10, Sarkar10, Adrados10,
  Gavrilov10-jetpl-en, Gavrilov12-prb}.  Thus, the very short lifetime
of polaritons ($\tau \sim 10^{-11}$\,s in GaAs-based microcavities)
makes them interesting in view of fast optical switches, but on the
other hand, it usually forbids long-lived transient or unsteady states
in a spatially homogeneous and truly constant environment.

Here we report a mechanism lifting the above limitation.  We consider
an internally homogeneous spinor polariton system pumped at normal
incidence and show that its well-known multistable behavior can be
turned into deterministic chaos.  This happens when a spontaneously
broken spin symmetry of the condensate (\cite{Gavrilov13-apl,Ohadi15})
starts being restored through Bogolyubov excitations, which
essentially relies on the non-Hermitian nature of polaritons and would
be impossible in ordinary Bose-Einstein condensates close to thermal
equilibrium.

Previously, both oscillatory~\cite{Sarchi08} and
chaotic~\cite{Solnyshkov09-j,Magnusson10} polariton states were
theoretically considered in a double-well geometry resembling a
Josephson junction.  The Josephson oscillations of polaritons were
observed under non-resonant excitation~\cite{Lagoudakis10,Abbarchi13}.
It is also reported that the combination of spatially separated
resonant and nonresonant pump sources is capable of producing a
chaotic spin state at a certain location~\cite{Ohadi15,Iorsh16}.
These mechanisms are limited to particular excitation geometries and
deliver only low-dimensional chaos.  In its turn, our mechanism
ensures a spatiotemporal spin chaos in arbitrarily sized polariton
systems, from pointlike micropillars to uniform planar cavities.

The following Section~\ref{sec:model} describes the model and the
instabilities leading to the chaotic behavior which is then analyzed
in the cases of pointlike systems (Sec.~\ref{sec:pointlike}) and large
cavities (Sec.~\ref{sec:space}).

\section{Spin symmetry breakdown}
\label{sec:model}

Coherent polariton states are described by the spinor Gross-Pitaevskii
equation~\cite{Kavokin-book-07},
\begin{equation}
  \label{eq:gpe}
  i \hbar \frac{\partial \psi_\pm}{\partial t} =
  \left(\hat E - i\gamma + V \psi_\pm^* \psi_\pm^{\vphantom *}\right)
  \psi_\pm^{\vphantom *}
  + \frac{g}{2} \psi_\mp^{\vphantom *}
  + f_\pm^{\vphantom *} e^{-i \frac{E_p}{\hbar} t}.
\end{equation}
A pair of macroscopic wavefunctions $\psi_\pm$ depend on time and
spatial coordinates in a two-dimensional active cavity layer.
$\psi_+$ and $\psi_-$ are the spin-up and spin-down components
connected directly with the right ($\sigma^+$) and left ($\sigma^-$)
circular polarizations of the output light.  $V$ is the
polariton-polariton interaction constant.  Polaritons with parallel
spins repel each other ($V > 0$)~\cite{Ciuti98, Renucci05,
  Vladimirova10, Sekretenko13-10ps}, which results in a blue shift
(${\sim} \, V |\psi^2|$) of their resonance energy.  Choosing $V = 1$
determines the units of $\psi$ and $f$.  Operator $\hat E = \hat E(-i
\hbar \nabla)$ implies the dispersion law common for both spin
components (see Appendix~\ref{app:model}); $\gamma$ is the decay rate.
The eigenstates at $\psi_\pm \to 0$ are polarized linearly in the $x$
and $y$ directions so long as $ \binom{\psi_x}{\psi_y} =
\frac{1}{\sqrt{2}} \left(
  \begin{smallmatrix}
    1 & 1 \\ i & -i
  \end{smallmatrix}
\right)
\binom{\psi_+}{\psi_-}
$
by definition, and $g \equiv E_x - E_y$ is the splitting between them.
The incident light wave (``pump'') has spinor amplitude $f_\pm$,
frequency $E_p / \hbar$, and a zero in-plane momentum ($k = 0$), so
that it falls perpendicular to the cavity.  Let us finally assume $f_+
= f_- = f$; then the pump is $x$-polarized and the equations for
$\psi_+$ and $\psi_-$ become exactly the same.  The plane-wave
solutions with $k = 0$ have the form $\psi(t) \propto e^{-i (E_p /
  \hbar) t}$ for both spin components.  Such solutions always exist in
a spatially homogeneous system, however, they can be unstable.

The exact spin symmetry ($\psi_+ = \psi_-$) can be spontaneously
broken if $g > 0$ and $E_p$ exceeds both $E_x$ and
$E_y$~\cite{Gavrilov13-apl}.  This is a purely dynamical effect,
unlike equilibrium phase transitions described by the Landau theory.
With increasing field density, the threshold is reached where both
$|\psi_+|$ and $|\psi_-|$ tend to jump sharply, which is attributed to
bistability~\cite{Gavrilov10-en}.  At the same time, the spin coupling
($\propto g$) makes one of the $\sigma^\pm$ components inhibit or feed
the other depending on their phase difference that, in turn, depends
on the amplitudes.  As a joint effect, it appears that near the
threshold an indefinitely small addition to one of $|\psi_+|$ or
$|\psi_-|$ triggers its further growth and the drop of the other
component, so that the condensate switches to the state with either
right or left circular polarization.

In the previous experiments \cite{Paraiso10, Gavrilov13-apl,
  Gavrilov13-mf, Sekretenko13-fluct, Gavrilov14-prb-j}, splitting $g$
was comparable to decay rate $\gamma$, and the condensate states with
broken spin symmetry were stable under constant excitation conditions.
Here we assume that $g$ exceeds $\gamma$ several times (e.\,g., due to
a lateral strain), which alters the system behavior drastically.  The
spin-asymmetric condensate becomes unstable with respect to the
scattering into two modes that are separated from it by finite energy
gaps $\pm \Delta E$ and have the same momentum $k = 0$.  The two
scattered modes are in fact two pairs of coupled same-energy yet
opposite-spin Bogolyubov excitations whose coupling and instability
are reachable only at sufficiently large~$g$.  This differs from the
known polariton-polariton scattering processes in which ``signal'' and
``idler'' have distinct momenta (e.\,g., \cite{Ciuti03, Carusotto04,
  Whittaker05, Solnyshkov08-dsp}).  More details and mode diagrams are
presented in Appendix~\ref{app:stability}.

Thus, both the spin-symmetric ($\Pi$) and asymmetric ($\Sigma_\pm$)
one-mode condensate states can be unstable.  The two instabilities
meet at, roughly, $g \gtrsim 4 \gamma$ and $2\gamma + g/2 \lesssim D
\lesssim 2g$, $D$ being the pump detuning from the unsplit polariton
level, $D \equiv E_p - E (k \, {=} \, 0)$.  When these conditions are
satisfied, no one-mode states are stable in a finite interval of~$f$,
irrespective of the dispersion law and mode structure.  This leads, on
one hand, to a chaotic behavior of the ``dispersionless'' polariton
condensates in cavity micropillars.  On the other hand, absence of
steady plane-wave solutions suggests spontaneous pattern formation in
spatially extended systems.  These two possibilities are analyzed
below.

\section{Evolution of pointlike states}
\label{sec:pointlike}

\begin{figure}
  \centering
  \includegraphics[width=\linewidth]{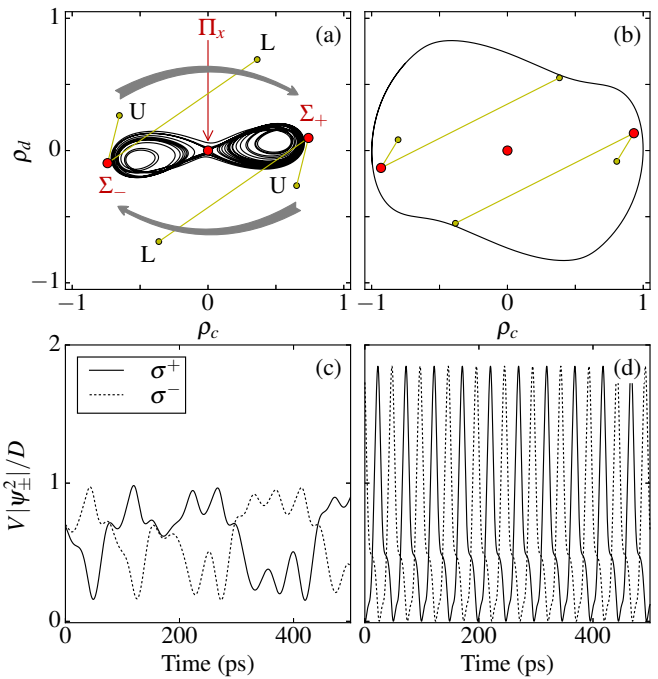}
  \caption{\label{fig:dynamics} (Color online) Dynamics of a pointlike
    polariton system.  Numerical solutions of Eq.~\eqref{eq:gpe} for
    $2f^2 = 5 \cdot 10^{-4}$ (a,~c) and $1.25 \cdot 10^{-3}$ (b,~d).
    (a,~b)~Phase trajectories over 2000~ps.  The large circles
    indicate the unstable one-mode condensate states ($\Pi_x$ and
    $\Sigma_\pm$).  The small circles show the respective lower (L)
    and upper (U) scattered modes.  (c,~d)~Explicit time dependences
    of $|\psi_\pm^2|$ over 500~ps.}
\end{figure}

Let us consider a pointlike system with $\gamma = 0.02$~meV (like in
modern GaAs-based cavities), $g = 10 \gamma$, and $D = 7.5 \gamma$.
Its evolution is described using only three degrees of freedom,
$|\psi_+|$, $|\psi_-|$, and phase difference $\arg \psi_+^*
\psi_-^{\vphantom *}$.  The pump is switched on smoothly and then kept
constant.  All the transient processes vanish in hundreds of
picoseconds and are not discussed here.  The pump has a small
stochastic component whose only purpose is to initiate the symmetry
breakdown at the very beginning; it can be arbitrarily small and does
not affect the conclusions.

To display phase trajectories [Fig.~\ref{fig:dynamics}(a), (b)], the
Stokes polarization parameters $\rho_c \equiv \rho_{(+,-)}$ and
$\rho_d \equiv \pldiag$ are chosen, which are the degrees of circular
and $\pm 45^\circ$ linear polarizations; here $\rho_{(a,b)} =
(|\psi_a^2| - |\psi_b^2|) / (|\psi_a^2| + |\psi_b^2|)$ by definition.
The $(x,y)$ polarization $\rho_l \equiv \pllin$ can be deduced as the
Stokes vector length is unity, $\rho_c^2 + \rho_l^2 + \rho_d^2 = 1$.
The positive (negative) sign of $\rho_d$ is indicative of the
$\sigma^- \rightarrow \sigma^+$ ($\sigma^- \leftarrow \sigma^+$)
conversion implied by the linear coupling term in Eq.~\eqref{eq:gpe};
to make sure, note that $g > 0$ and $\sgn [\Im (\psi_+^*
\psi_-^{\vphantom *})] = \sgn [\rho_d (\psi)]$ for each spinor $\psi$.
In line with this rule, $\sgn (\rho_c) = \sgn (\rho_d)$ in each of the
$\Sigma_\pm$ condensate states.  However, the scattered modes appear
to have the opposite sign of $\rho_d$, and thus their filling acts to
restore the spin symmetry or even flip the spin, provided they are
filled rapidly enough.

The dynamics of a chaotic polariton state [Fig.~\ref{fig:dynamics}(a),
(c)] can be understood as a counteraction between the symmetry
breakdown tendency and the instability of the asymmetric states.  The
two instabilities differ in nature, therefore they cannot balance each
other.  At lower $f$ (not shown) the system swings between the $\Pi$
state and one of the $\Sigma_\pm$ states but does not alter the
initially ``chosen'' sign of $\rho_c$.  With increasing $f$, the
growth rate $\Gamma$ of the scattered modes increases, and so the
condensate is able to occasionally flip the spin.  Instead of the
single---spontaneous and irreversible---symmetry breakdown, one
observes an infinite sequence of spin switches; nevertheless, the
system is still extremely sensitive to fluctuations near the $\Pi$
point.  Calculations show that an arbitrarily small deflection from
any given point of the phase trajectory results in an exponentially
divergent trajectory, which is indicative of chaos.

At a stronger pump, $\Gamma$ becomes so large that it ensures the
spin-flip event each time the system reaches a $\Sigma$ state, and the
trajectory takes the form of a limit cycle
[Fig.~\ref{fig:dynamics}(b), (d)].  Maximum $\Gamma$ is roughly
comparable to decay rate $\gamma$ (e.\,g.,
\cite{Ciuti03,Gavrilov14-prb-b}); accordingly, the oscillation period
is comparable to the polariton lifetime.  With further increasing $f$
(not shown), $\Gamma$ is reduced, as the coupling of the Bogolyubov
excitations decays due to their energy mismatch (this is explained in
detail in Appendix~\ref{app:stability}).  As a result, chaotic
dynamics re-establishes; afterwards it degenerates into one-scroll
oscillations and, finally, into a fixed point.

\begin{figure}
  \centering
  \includegraphics[width=0.8\linewidth]{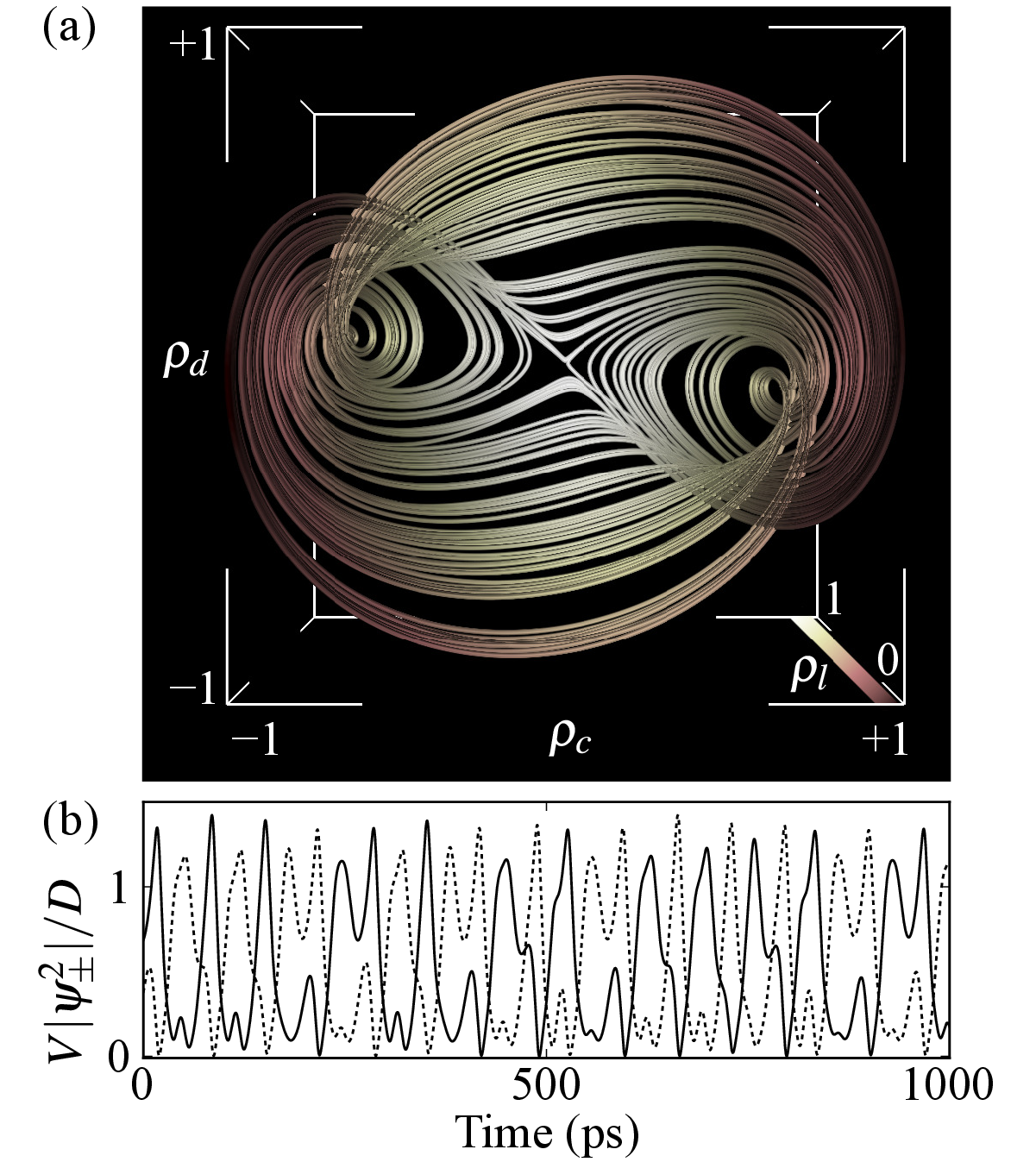}
  \caption{\label{fig:attractor} (Color online) Chaotic attractor
    typical of a pointlike system.  (a)~Phase trajectory over a
    6000~ps interval in the space of the Stokes polarization
    parameters; the color represents linear polarization $\rho_l$.
    (b)~Explicit time dependences of $|\psi_\pm^2|$ over 1000~ps.}
\end{figure}

With increasing pump detuning up to $D = g$, the nonlinearity is
enhanced and the system behavior becomes chaotic in a wider interval
of~$f$.  Figure~\ref{fig:attractor} shows the evolution at $D = g = 10
\gamma = 0.2$~meV and $2f^2 = 5 \cdot 10^{-4}$.  The spin flips are
quick and irregular and it is impossible to predict the dominant spin
for more or less far future.  Even when absolutely no stochastic
factors influence the system, prediction of its future states requires
an exponentially growing precision.  Similar features implemented in
lasers were proposed for fast random number
generation~\cite{Sciamanna15}.

\section{Self-organization and chaos in planar cavities}
\label{sec:space}

\begin{figure}
  \centering
  \includegraphics[width=0.8\linewidth]{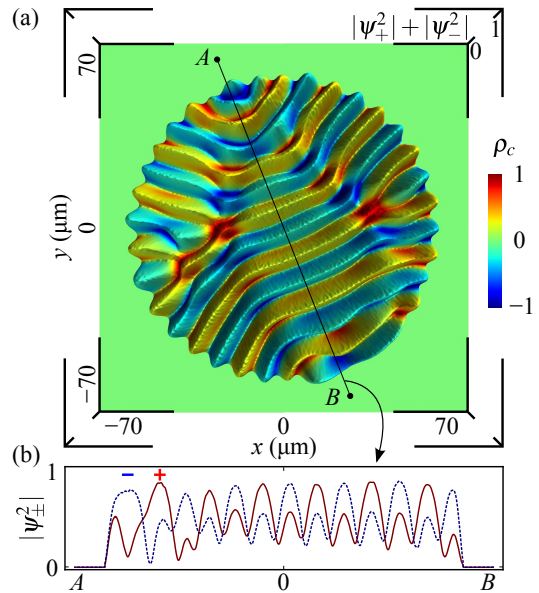}
  \caption{\label{fig:wires} (Color online) Self-organized field in a
    circular cavity. (a)~Full intensity $|\psi_+^2| + |\psi_-^2|$, in
    arbitrary units, as a function of in-plane coordinates; the color
    represents circular polarization $\rho_c$.  (b)~Explicit spatial
    dependences of $|\psi_\pm^2|$ along the cross section~$AB$.}
\end{figure}

Let us now turn to spatially extended polariton systems under one-mode
driving.  \emph{A priory} it is clear that if such a system keeps
being plane-wave, it will fit the already discussed scenario
(spatially homogeneous solutions must be rapidly varying), otherwise
spatial symmetry will be broken.

To ensure zero boundary conditions, the decay rate $\gamma$ is set to
increase sharply on the edge of a circular area of radius $R =
60~\mu$m.  Hence, the model is $\mathrm{SO}(2)$-invariant, i.\,e.,
rotationally symmetric.  Let us note that the anisotropy implied by
both the splitting $(g \equiv E_x - E_y)$ and $x$-oriented pump
polarization is reflected only in phase difference $\arg \psi_+^*
\psi_-^{\vphantom *}$.  At the same time, $\psi_+(x, y)$ and
$\psi_-(x, y)$ per se are expected to be rotationally symmetric, based
on the symmetry of the potential landscape and excitation conditions.

Figure~\ref{fig:wires} displays the cavity-field distribution obtained
at $g = 10 \gamma$ and $D = 7.5 \gamma$ (as in
Fig.~\ref{fig:dynamics}).  In this example, the system comes to a
steady but highly nonuniform state in which the continuous
$\mathrm{SO}(2)$ symmetry is broken together with the spin symmetry.
Such solutions are established in hundreds of picoseconds and are
spontaneous in that a small fluctuation coming at a critical stage
would result in a completely different picture.  They are
self-organized, i.\,e., ordered internally, and always ``preferred''
over rapidly varying homogeneous solutions if it is ever allowed
dynamically.

The scale of inhomogeneity is deduced from the wave numbers $k_s$ of
the scattered modes.  The parametric scattering \emph{from} the
condensate always precedes and mediates its switches to high-energy
states in bi- or multistable polariton
systems~\cite{Gavrilov14-prb-b,Gavrilov15}.  In accordance with the
dispersion law and phase-matching conditions, in our system $|k_s|$
may vary from 0 to ${\sim} \, 0.4~\mu\mathrm m^{-1}$ depending on
$\psi_\pm$ (see Appendix~\ref{app:stability}).  Correspondingly, the
cavity field is inhomogeneous on the scale of several microns
irrespective of~$R$.

The very intriguing feature of the self-organized cavity field is that
it arranges itself into a bunch of wires that have alternating
$\rho_c$ and are separated by low-intensity regions. This effect
originates in the free flow of polaritons combined with their
repulsive interaction, which becomes especially pronounced in
spatially distributed bistable systems.  For instance, if the switch
to the ``on'' state in such a system proceeds locally, the polariton
flow triggers a chain of analogous switches in neighboring places, and
eventually the ``on'' state with a particular spin spreads as far as
possible~\cite{Liew08-prl-neur,Sekretenko13-fluct}.  In our system,
the polariton flow acts to expand and, thus, homogenize both the
spin-up and spin-down high-density regions.  As a result, given that
flat and ``rapid'' two-dimensional patterns are ruled out in favor of
self-organized steady states, the only possible outcome of the spin
expansion is a set of long-ordered quasi-one-dimensional wires.  The
\emph{filamentation} is in fact typical of many of spontaneously
ordered systems~\cite{Cross93,Bohr98}.

\begin{figure}
  \centering
  \includegraphics[width=1\linewidth]{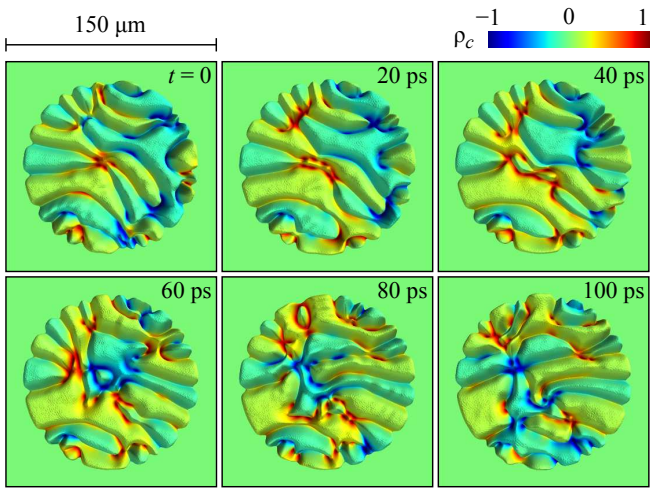}
  \caption{\label{fig:turbulence} (Color online) Chaotic field
    evolution.  Instantaneous emission snapshots drawn at 20~ps
    intervals.  The color represents circular polarization~$\rho_c$.}
\end{figure}

The spin wires are inevitably destroyed at larger field densities.
They always tend to expand in all directions, and so the chaotic
dynamics is restored as soon as sufficiently broad homogeneous
patterns return in the field distribution.  Such patterns are
comparatively narrow in the $k$ space and, hence, internally unstable
due to the same reason as the one-mode solutions considered above in
Secs.~\ref{sec:model} and \ref{sec:pointlike}.
Figure~\ref{fig:turbulence} shows the evolution over a 0.1~ns interval
starting at 2.5~ns after the constant pump has been switched on; it is
obtained at $g = 15 \gamma$ and $D = 12.5 \gamma$.  The field
structure reveals spatiotemporal chaos.  It is highly sensitive to
fluctuations at all times rather than only at a certain critical stage
of pattern formation.  At every instant, an arbitrary small and
short-range fluctuation would lead to an exponentially divergent
evolution branch.  As a result, each given location within the cavity
plane shows the chaotic behavior.

Thus, beyond a certain critical scale $\mathfrak L$ the cavity field
cannot be at once steady and spatially homogeneous; furthermore, the
inhomogeneous steady states lose stability at higher field densities.
An example of a limit-cycle solution with a spontaneous reduction of
the $\mathrm{SO}(2)$ symmetry down to $\mathrm{C}_2$ for $R = 10~\mu$m
is analyzed in Appendix~\ref{app:periodic}; it is an intermediate
stage between pointlike states and large two-dimensional cavities.
With increasing size, closed periodical trajectories cease to exist.
It is reasonable that the degree of complexity of such systems
infinitely grows with $R$, in analogy to turbulent
fluids~\cite{Bohr98}.  This hypothesis is substantiated, first, by
perfectly deterministic character of the model, i.\,e., the mere fact
that complexity does not come from the outer sources of entropy like
thermal reservoirs, random potentials, or stochastic forces.  On the
other hand, enlarging extent does enrich at least a statistical
diversity.  Numerical estimates show that the first-order spatial
correlation function drops rapidly with increasing distance beyond
$\mathfrak L$, which suggests statistical independence of remote
locations.  Nonetheless, the geometry of the high-dimensional chaotic
attractors in cavity-polariton systems invites further investigation.

\section{Conclusion}
\label{sec:conclusion}

In summary, we have found the mechanism making spin of light behave as
a turbulent fluid.  The key point is the spontaneous breakdown of spin
symmetry, which brings about an extreme sensitivity to fluctuations in
a resonantly excited polariton system.  The scattering into Bogolyubov
modes acts to close the cycle and triggers an infinite series of spin
switches.  This mechanism is fundamental; being independent of the
potential landscape or the shape of the pump wave, it is expected to
take place in arbitrarily sized planar microcavities.  As a result,
cavities with strong exciton-photon coupling can serve as all-optical
free-running chaotic radiation emitters operating on the scale of
picoseconds and microns.

\begin{acknowledgements}
  I wish to thank N.~A.~Gippius, S.~G.~Tikhodeev, and
  V.~D.~Kulakovskii for stimulating discussions.  The work was
  supported by the Russian Science Foundation (grant No.\
  14-12-01372).
\end{acknowledgements}

\appendix

\section{Two-dimensional model}
\label{app:model}

The low-polariton dispersion law reads,
\begin{equation}
  E_\mathrm{LP}(\mathbf k) = \frac{E_C(\mathbf k) + E_X}{2}
  - \frac12 \sqrt{[ E_C(\mathbf k) - E_X ]^2 + \mathfrak R^2},
\end{equation}
where $\mathfrak R$ is the exciton-photon coupling rate (Rabi
splitting), $E_X$ the exciton energy, and $E_C(\mathbf k)$ the
dispersion law of cavity photons,
\begin{equation}
  E_C^{\mathstrut}(\mathbf k) =
  E_C^{(0)} + \frac{\hbar^2 \mathbf k^2}{2 m_C},
  \qquad m_C = \frac{\varepsilon E_C^{(0)}}{c^2}.
\end{equation}
The exciton effective mass is much larger than $m_C$, so the $k$
dependence of $E_X$ is neglected.  In the real space, energy operator
$\hat E$ takes the form
\begin{equation}
  \hat E = E_\mathrm{LP}(-i \hbar \nabla) +
  \begin{cases}
    0&\!\!\!\text{at $|\mathbf r| \le R$;}\\
    U&\!\!\!\text{at $|\mathbf r| > R$.}
  \end{cases}
\end{equation}
This is the bare---unperturbed and unsplit---polariton energy
substituted into Eq.~\eqref{eq:gpe}.  The jump at $|\mathbf r| = R$
enables simulating a finite-sized cavity; it should be just large
enough to ensure zero boundary conditions.  For instance,
qualitatively the same results are obtained for $U = 25$~meV, $U =
-25$~meV, and $U = -i \times 25~\mathrm{meV}$.  The last variant means
a jump in the decay rate; it was used in final simulations.  The
microcavity parameters are $E_C^{(0)} = E_X^{\vphantom)} = 1.5$~eV,
$\varepsilon = 12.5$, and $\mathfrak R = 10$~meV.  Eq.~\eqref{eq:gpe}
was solved on a $250 \times 250$ square grid with $-75 < x, y <
75~\mu$m.

\section{Plane-wave solutions and their asymptotic stability}
\label{app:stability}

\begin{figure}
  \centering
  \includegraphics[width=\linewidth]{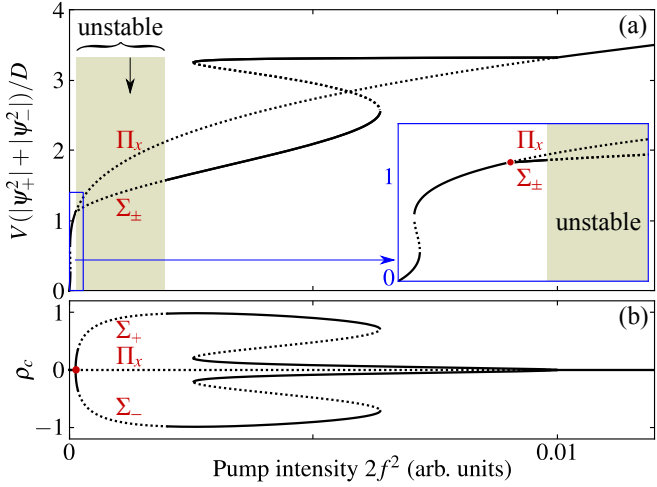}
  \caption{\label{fig:onemode} (Color online) (a) One-mode solutions
    of Eq.~\eqref{eq:onemode}, condensate intensity vs.\ pump
    intensity.  The range of small pump intensities is shown in the
    inset.  Unstable branches are indicated by dotted lines.  The
    interval with no stable solutions is shadowed.  (b)~Corresponding
    degrees of circular polarization.}
\end{figure}

Since the pump is a plane wave falling perpendicular to the cavity ($k
= 0$), the stationary solutions of Eq.~\eqref{eq:gpe} are sought for
in the form $\psi_\pm(t) = \bar\psi_\pm e^{-i (E_p / \hbar) t}$.  One
obtains the following time-independent equations for $\bar\psi_+$ and
$\bar\psi_-$:
\begin{equation}
  \label{eq:onemode}
  \left[
    E_p
    - \left( E_\mathrm{LP}(k \, {=} \, 0) - i \gamma \right)
    - V \bar\psi_\pm^* \bar\psi_\pm^{\vphantom *}
  \right] \bar\psi_\pm^{\vphantom *}
  - \frac{g}{2} \bar\psi_\mp^{\vphantom *} - f = 0.
\end{equation}
Stability analysis is performed by linearizing Eq.~\eqref{eq:gpe} over
small deflections from the one-mode solutions.  In the $k$ space this
leads to the $4 \times 4$ linear problem
\begin{gather}
  \label{eq:lin}
  \hat L \Psi = E \Psi, \quad \text{where}\\
  \Psi =
  \left[
    \psi_+^{\vphantom *} (\mathbf k),
    \psi_+^* (-\mathbf k),
    \psi_-^{\vphantom *} (\mathbf k),
    \psi_-^*(-\mathbf k)
  \right]^\mathrm{T},
\end{gather}
\begin{widetext}
  \begin{equation}
    \label{eq:matrix}
    \hat L =
    \begin{pmatrix}
      E_\mathrm{LP}(\mathbf k) - i\gamma + 2V |\bar\psi_+^2|
      & V \bar\psi_+^2
      & g/2
      & 0
      \\ -V \bar\psi_+^{*2}
      & 2E_p - E_\mathrm{LP}(-\mathbf k) + i\gamma - 2V |\bar\psi_+^2|
      & 0
      & -g/2
      \\ g/2
      & 0
      & E_\mathrm{LP}(\mathbf k) - i\gamma + 2V |\bar\psi_-^2|
      & V \bar\psi_-^2
      \\ 0
      & -g/2
      & -V \bar\psi_-^{*2}
      & 2E_p - E_\mathrm{LP}(-\mathbf k) + i\gamma - 2V |\bar\psi_-^2|
    \end{pmatrix}.
  \end{equation}
\end{widetext}

Below we take $\gamma = 0.02$~meV, $g \equiv E_x - E_y = 10 \gamma$,
and $D \equiv E_p - E_\mathrm{LP}(k \, {=} \, 0) = 7.5 \gamma$ (as in
Figs.~\ref{fig:dynamics} and \ref{fig:wires}).  The one-mode solutions
$\bar\psi_\pm$ are determined using Eq.~\eqref{eq:onemode} in a wide
range of $f$ (Fig.~\ref{fig:onemode}).  Solving the linear
problem~\eqref{eq:lin} for each $\bar\psi_\pm$ yields, in each of two
spin components, a pair of Bogolyubov excitations obeying
phase-matching conditions
\begin{equation}
  \label{eq:phase_matching}
  E_\mathrm{signal}(\mathbf k) + E_\mathrm{idler}(-\mathbf k) = 2E_p.
\end{equation}
``Signal'' and ``idler'' are widely used designations for such modes
in cavity-polariton systems, by analogy with optical parametric
oscillators.  Eq.~\eqref{eq:phase_matching} means energy and momentum
conservation in the course of the two-particle scattering from the
pumped mode, $(p, p) \to (s, i)$. The real and imaginary parts of the
eigenvalues of matrix~\eqref{eq:matrix} determine, respectively, the
energy levels and decay rates of the excitations; a positive imaginary
part means instability.  The Stokes polarization parameters are
determined by the respective eigenvectors.  Figure~\ref{fig:onemode}
intentionally does not take account of the scattering into $k \neq 0$;
in other words, it shows only the one-mode (in)stability.

The range of low pump amplitudes is shown in the inset of
Fig.~\ref{fig:onemode}(a).  The pump polarization ($x$) matches
exactly the upper eigenstate at $\bar\psi_\pm \to 0$ which is, hence,
the only occupied state at small $f$.  At the very beginning the
response is linear and afterwards it takes the form of an S-shaped
curve.  Since $E_p - E_x$ exceeds $\sqrt{3} \gamma$, the system is
\emph{bistable}~\cite{Elesin73,Baas04-pra,Gavrilov14-prb-b}, which
stems from the positive feedback loop between the amplitude
($|\psi_x|$) and effective resonance frequency ($E_x + V|\psi_x|^2$)
in a finite range of $|\psi_x|$.

\begin{figure}
  \centering
  \includegraphics[width=\linewidth]{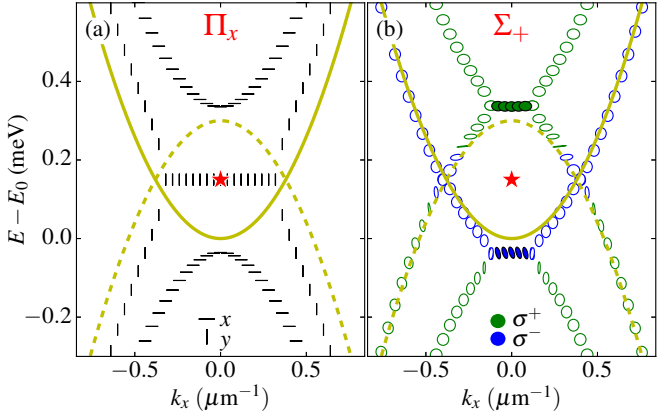}
  \caption{\label{fig:dispersion} (Color online) Bogolyubov
    excitations in the $\Pi_x$~(a) and $\Sigma_+$~(b) one-mode states
    at $2f^2 = 1.25 \cdot 10^{-3}$ [the pump intensity is marked by
    the vertical arrow at the top of Fig.~\ref{fig:onemode}(a)].
    Stars indicate the condensate mode at $k = 0$, $E = E_p$.  Lines
    show the reference (unsplit) dispersion law $E_\mathrm{LP}(k)$
    (solid line) and its ``idler'' counterpart $2E_p -
    E_\mathrm{LP}(-k)$ (dashed line).  Strips and ellipses represent
    polarization states.  Full ellipses in (b) indicate unstable
    modes.}
\end{figure}

With increasing $f$, the nonlinear terms in Eq.~\eqref{eq:gpe} couple
the $x$ and $y$ polarization components.  As a result, one-mode
solutions with a nonzero fraction of the ``unpumped'' ($y$) component
become possible along with purely $x$-polarized states.  That is why
the upper stability branch splits into three, of which one ($\Pi_x$)
remains polarized linearly in the $x$ direction and the other two
($\Sigma_\pm$) acquire large and mutually opposite circular
polarizations.  All these one-mode branches are unstable over a wide
interval of $f$.  The excitations in the $\Pi_x$ and $\Sigma_+$ states
near the center of that interval are outlined in
Fig.~\ref{fig:dispersion}.

\textit{Branch $\Pi$ and the symmetry
  breakdown.}---Figure~\ref{fig:dispersion}(a) reveals the
$y$-polarized ``signal'' and ``idler'' stuck together within a wide
flat area centered at $E = E_p$.  Since the real parts of their
energies coincide, the imaginary parts diverge and one of them takes a
positive value; this is quite a general property of
Eq.~\eqref{eq:gpe}~\cite{Ciuti03}.  Thus, on branch $\Pi_x$ the
condensate is unstable \emph{in itself}: its $y$-polarization
component tends to jump sharply.  Consequently, the field has to
acquire an elliptical polarization, but Fig.~\ref{fig:dispersion}(a)
suggests nothing about its \emph{sense}; the spin-up ($\Sigma_+$) and
spin-down ($\Sigma_-$) condensate states that could arise as a result
of such instability are in fact equally
probable~\cite{Gavrilov13-apl}.

\begin{figure}
  \centering
  \includegraphics[width=0.85\linewidth]{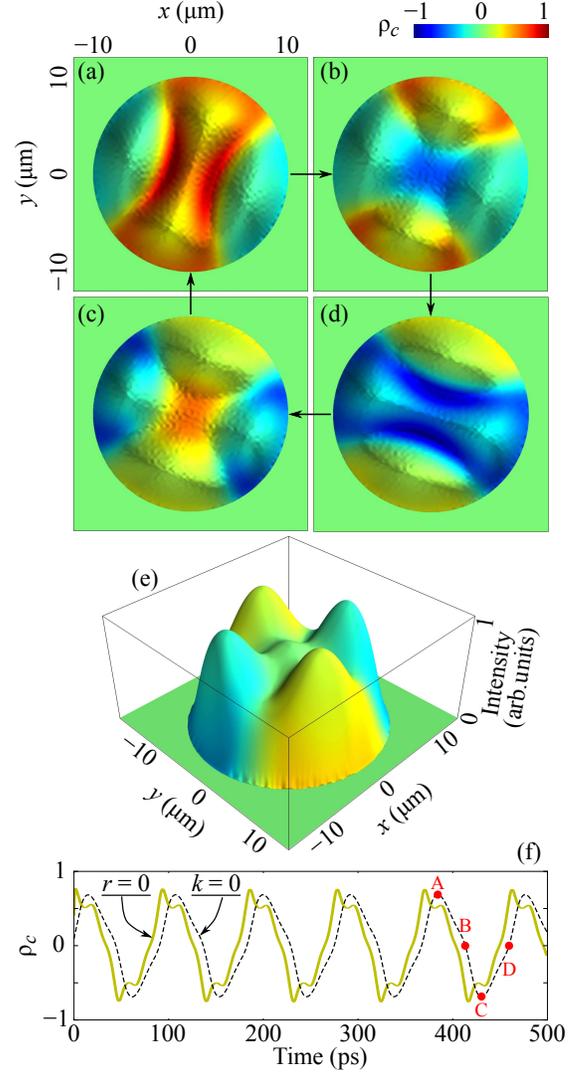}
  \caption{\label{fig:disk_small} (Color online) Periodic solution for
    $R = 10~\mu$m.  (a)--(d)~Instantaneous spin and intensity
    patterns; (e) the pattern averaged over the period.  (f) Time
    dependences of $\rho_c$ in the $\mathbf r = 0$ point of the real
    space (solid line) and $\mathbf k = 0$ point of the momentum space
    (dashed line).  The time moments represented in panels (a)--(d)
    are marked as A--D.}
\end{figure}

\textit{Branch $\Sigma$ and coupled Bogolyubov modes.}---The $\Sigma$
states of the condensate are stable at comparatively small $g$.  The
excitations at $k = 0$ were previously known to be occupied in the
course of transient processes but not in a steady state, as they were
decaying~\cite{Brichkin15}.  Increasing $g$ brings about unstable
(non-decaying) modes below and above the condensate as seen in
Fig.~\ref{fig:dispersion}(b).

The bottom row of unstable modes is found near the weakly populated
and, thus, unshifted level of the minor spin component ($\sigma^-$ in
our example).  Approximately the same energy corresponds to a
significantly red-shifted $\sigma^+$-polarized ``idler'', if we accept
that conventional idlers form the downward branches.  The
opposite-spin signals and idlers with nearly coincident energies stick
together and so they make up the flattened $k$-space region and
acquire positive growth rates $\Gamma \equiv \Im E$.  As said above,
it is possible only at comparatively large opposite-spin coupling~$g$.

The greater the amplitude, the stronger the blue (red) shift of the
signal (idler).  With increasing $f$, the major-spin idler goes far
below $E_0 = E_\mathrm{LP}(k \, {=} \, 0)$ and is no longer able to
hybridize with the other---always weak and unshifted---spin component
because of their energy mismatch.  As a result, the one-mode
$\Sigma_\pm$ states cannot break up anymore and, thus, restore their
stability.  In general, there is the other way to restore stability at
larger $f$ [Fig.~\ref{fig:onemode}] via a jump in \emph{both}
$\sigma^\pm$ components in spite of the symmetry breaking forces;
this, however, does not happen in the considered parameter range.

\section{Periodic states in small cavities}
\label{app:periodic}

Let us consider an intermediate case between pointlike states and
large planar systems.  The solution sketched out in
Fig.~\ref{fig:disk_small} is obtained for a small circular cavity with
$R = 10~\mu$m.  The parameters are $D = g = 10 \gamma = 0.2$~meV and
$2f^2(k\,{=}\,0) = 3 \cdot 10^{-3}$.  They lead to the area with no
steady-state solutions; in the outer areas the spin and intensity
distributions are always constant in time and rotationally invariant.

Figure~\ref{fig:disk_small} shows that (i) the $\mathrm{SO}(2)$
symmetry is spontaneously reduced down to $\mathrm C_2$, (ii) the spin
symmetry is broken, and (iii) the space pattern oscillates with a
period of $T \approx 90$~ps.  The distinctive stages of the cycle are
shown in Figs.~\ref{fig:disk_small}(a)--(d).  The field is anisotropic
even after averaging over the period [Fig.~\ref{fig:disk_small}(e)].
It exhibits a pair of orthogonal directions that are spontaneously
chosen at the moment of symmetry breaking.

Figure~\ref{fig:disk_small}(f) shows the dynamics of the
circular-polarization degree in the $\mathbf r = 0$ point of the real
space and in the $\mathbf k = 0$ point of the momentum space.  Both
have the amplitude of about 0.7.  The integral polarization,
\begin{equation*}
  \bar\rho_c = \frac{I_+ - I_-}{I_+ + I_-},
  \quad \text{where} \quad
  I_\pm(t) = \int |\psi_\pm(\mathbf r, t)|^2 d \mathbf{r},
\end{equation*}
has a smaller oscillation amplitude of 0.25 (not shown); the full
intensity $I_+(t) + I_-(t)$ also varies with time within about 10\%.
Thus, the evolution is reduced to neither plain temporal oscillations
typical of micropillars nor a periodic spatial redistribution with
invariant integral characteristics that would be rather typical of the
Josephson effect in Hermitian systems.

The island-like patterns grow into the spin wires with increasing~$R$.
On the other hand, increasing density at greater $g$ and $D \sim g$
would homogenize the system, thereby making it similar to plain
pointlike states in small cavity micropillars.


%

\end{document}